\documentclass{IEEEtran}
\usepackage{graphicx}
\usepackage{algorithm}
\usepackage{algorithmic}
\usepackage{amsmath, amsthm} 
\usepackage{amsfonts}
\usepackage{cite}
\usepackage{booktabs}

\newtheorem{definition}{Definition}
\newtheorem{theorem}{Theorem}
\title{Learning Sheaf Laplacian Optimizing \\Restriction Maps}
\author{Leonardo Di Nino$^{1,2}$,  Sergio Barbarossa$^1$ and Paolo Di Lorenzo$^{1,2}$\medskip\\
    $^1$ Department of Information Engineering, Electronics, and Telecommunications, Sapienza University, Rome, Italy \smallskip\\
    $^2$ National Interuniversity Telecommunications Consortium (CNIT), Parma, Italy \smallskip\\    
    Emails: \{leonardo.dinino,sergio.barbarossa,paolo.dilorenzo\}@uniroma1.it
    \thanks{This work was supported by Huawei Technology France SASU under Grant TC20220919044. The work of P. Di Lorenzo has been supported by the SNS JU project 6G-GOALS \cite{strinati2024goal} under the EU’s Horizon program Grant Agreement No 101139232.}\vspace{-.2cm}}
\begin{document}
\maketitle

\begin{abstract}
The aim of this paper is to propose a novel framework to infer the sheaf Laplacian, including the topology of a graph and the restriction maps, from a set of data observed over the nodes of a graph. The proposed method is based on sheaf theory, which represents an important generalization of graph signal processing.  
The learning problem aims to find the sheaf Laplacian that minimizes the total variation of the observed data, where the variation over each edge is also locally minimized by optimizing the associated restriction maps. 
Compared to alternative methods based on semidefinite programming, our solution is significantly more numerically efficient, as all its fundamental steps are resolved in closed form. 
The method is numerically tested on data consisting of vectors defined over subspaces of varying dimensions at each node. We demonstrate how the resulting graph is influenced by two key factors: the cross-correlation and the dimensionality difference of the data residing on the graph's nodes.
\end{abstract}

\section{Introduction}
Graph representation learning and graph signal processing (GSP) have gained significant attention in recent years due to their ability to effectively model and interpret pairwise relationships inherent in structured data \cite{dong2020graph}. The generalization to higher order combinatorial structures, such as simplicial and cell complexes was also proposed in  \cite{Barbarossa_2020} and \cite{sardellitti2024topological}, to account for higher order interactions and handle signals defined over sets of any order (not only nodes of a graph). Representing data over non-Euclidean spaces has become particularly relevant in machine learning, where a fundamental assumption is that the input high-dimensional data typically lie on an intrinsically low-dimensional Riemannian manifold. A principled way to extract information from the data passes then through the estimation of the manifold, as suggested in \cite{lin2008riemannian}. More recently, the design of convolutional filters accounting for the manifold's geometric structure was propposed in  \cite{battiloro2024tangentbundleconvolutionallearning}.  Prototypical examples of problems include data alignment and synchronization \cite{bandeira2013multireferencealignmentusingsemidefinite}, where the decisional problem of establishing relations is deeply intertwined with the operational mechanism of linking locally defined observations in a meaningful way.

Notwithstanding the success of GSP, the field can still have a fundamental boost by the application of sheaf theory to signal processing and learning, as already proposed in \cite{robinson2014topological}.
Sheaf theory provides a powerful and flexible tool for modeling data defined on topological spaces and living on very general spaces, such as vector spaces,  groups or sets \cite{maclane2012sheaves}. In a nutshell, sheaf theory builds on two fundamental building blocks: the definition of a topological space, as a formal way to capture neighborhood relations, and the assignment of a space to each element (open set) of the topological space, satisfying some kind of continuity relation between neighbor elements, in a very general form. Within sheaf theory, the sheaf Laplacian operator extends the classical notion of the discrete Hodge  Laplacian defined over cell complexes and it represents a key operator to extract global properties from local relations. The inference of the sheaf Laplacian from empirical data is then a crucial step in applying sheaf-theoretic tools to practical problems.
Cellular sheaves not only address these problems effectively but also encompass and generalize graph and topological signal processing as specific cases.

\textbf{Motivation and Related Works}. We focus on cellular sheaves valued on vector spaces, to exploit the rich algebraic structure that can be associated with cell complexes. Specifically, cellular sheaves consist in the assignment of a vector space to each cell (nodes and edges in the graph case) and the definition of relationships between these vector spaces expressed as linear transformations. Within this framework, the sheaf Laplacian encapsulates both the geometric and topological perspectives. This extension provides the foundation for a spectral theory that underpins many applications of cellular sheaves. Key contributions in this area include modeling opinion dynamics in social networks \cite{hansen2020opiniondynamicsdiscoursesheaves}, enabling distributed optimization through homological programming \cite{8919796}, knowledge graph representation learning \cite{gebhart2023knowledgesheavessheaftheoreticframework} and designing advanced deep neural architectures that generalize message-passing networks \cite{hansen2020sheafneuralnetworks}. Our work focuses on learning the sheaf Laplacian and the most relevant paper in this area is \cite{8683709}, where the authors considered the problem of learning the sheaf laplacian from a set of data observed over the nodes of a graph, in order to minimize the total variation, generalizing the graph-based inferential procedure of \cite{pmlr-v51-kalofolias16} to cellular sheaves. 

\textbf{Contributions.} In this work, hinging on the algebraic structure of a graph cellular sheaf, we propose an algorithm to learn, jointly, the topology of the graph concurrently with  the restriction maps associated to each edge, up to a rotation matrix,  in order to minimize the total variation over the graph. The method extends the approach proposed in \cite{chepuri2016learningsparsegraphssmoothness}, incorporating the optimization of the restriction maps that minimize the variation along each edge. We start extracting a compact representation of the data and then we infer the structure of the graph,  using a generalized notion of distance between the spaces living on the nodes of the graph, optimizing the restriction maps on each edge, up to a rotation matrix. We show how this distance comes to depend only on the correlation between data living on different nodes and on the dimension of the vector spaces, rather than on their specific structure. The method is numerically efficient because all the basic steps are obtained in closed form. 
We assess our proposed strategy on synthetic data to check the validity of our approach, showing the potentials with respect to standard graph-based methods that do not optimize over the restriction maps.
\section{Background}
In this section, we review some basic notions of cellular sheaves theory. We assume that the data are defined over a cell complex. The general definition for a cellular sheaf involves the introduction of a face incidence poset $P_X$ (the partial ordered set of inclusion relations between cells) of a regular cell complex $X$ and a category forming the spaces of data to be assigned to $X$. The formal definition is \cite{maclane2012sheaves}:
\begin{definition}[Cellular Sheaves]
    A \textit{cellular sheaf} valued in a category $\mathcal{C}$ on a regular cell complex $X$ is a covariant functor $\mathcal{F}: P_X \rightarrow \mathcal{C}$.
\end{definition}

In particular, we consider the category $\mathcal{C} =$ \textbf{Vect} of vector spaces, such that the sheaf as a functor is specficied by: 

\begin{itemize}
    \item A vector space $\mathcal{F}(\sigma), \forall \sigma \in X$,
    \item A morphism $\mathcal{F}_{\sigma \triangleleft \tau}: \mathcal{F}(\sigma) \rightarrow \mathcal{F}(\tau)$ for each incidence $\sigma \triangleleft \tau$ (which for $\mathcal{C} =$ \textbf{Vect} is a linear transformation) satisfying the functiorality condition that if $\sigma \triangleleft \tau \triangleleft \xi$, then $\mathcal{F}_{\sigma \triangleleft \xi} = \mathcal{F}_{\sigma \triangleleft \tau} \circ \mathcal{F}_{\tau \triangleleft \xi}$
\end{itemize}

Since a graph is a specific case of a regular cell complex where the only incidence relation is the node-edge one, the definition further simplifies into: 
\begin{definition}[Sheaves on Graphs]
    Given a graph $G(V,E)$, a \textit{cellular sheaf} of vector spaces on the graph $(G, \mathcal{F})$ consists of:
    \begin{enumerate}
        \item A vector space $\mathcal{F}(v)$ for each $v \in V$;
        \item A vector space $\mathcal{F}(e)$ for each $e \in E$;
        \item A linear map $\mathcal{F}_{v \triangleleft e}:\mathcal{F}(v) \rightarrow \mathcal{F}(e)$ for each attachment $v \triangleleft e$ 
        of a higher dimensional cell (edge) $e$ to a lower dimensional cell (vertex) $v$, called a \textit{restriction map}
    \end{enumerate}
\end{definition}

The vector spaces defined over each node $\mathcal{F}(v)$ and over each edge $\mathcal{F}(e)$ are called \textit{stalks}.
    Two important vector spaces can be defined via direct sum of the stalks defined over nodes and edges respectively:
        \begin{itemize}
            \item The space of \textit{0-cochains} $C^0(G,\mathcal{F}) = \bigoplus_{v \in V} \mathcal{F}(v)$
            \item The space of \textit{1-cochains} $C^1(G,\mathcal{F}) = \bigoplus_{e \in E} \mathcal{F}(e)$
        \end{itemize}
    When considering a vector-valued sheaf, we may view a vector $\mathbf{x} \in C^0(G, \mathcal{F})$ as a concatenation of all the vectors locally observed over nodes: for the purposes of our work, we define our procedure starting from globally observed data belonging to $C^0(G, \mathcal{F})$.
    
    The \textit{global sections} space $H^0(G, \mathcal{F})$ is a linear subspace of $C^0(G, \mathcal{F})$ consisting of all the 0-cochains satisfying local consistency relationship for each incidency relation: it is the kernel of the \textit{coboundary map} $\delta:C^0(G,\mathcal{F}) \rightarrow C^1(G,\mathcal{F})$, being a linear operator acting block-wise on the 0-cochains as follows: 
        \begin{equation} 
        \label{coboundary}
            \delta(\mathbf{x})_e = \mathcal{F}_{u \triangleleft e}\mathbf{x}_u - \mathcal{F}_{v \triangleleft e}\mathbf{x}_v
        \end{equation}
Similarly to how we derive the graph Laplacian from the incidence matrix, we can define the sheaf incidence and the sheaf Laplacian from the coboundary map. As with graphs, to define the sheaf incidence, we need to define an orientation for each edge. Without loss of generality, denoting by $V$ and $E$ the number of nodes and edges in the graph, respectively, and assuming the stalks associated to each node and each edge to have dimension $d$, the sheaf incidence matrix is a block matrix of dimension $d V \times d E$, whose blocks are $d \times d$ and are built as follows:
\begin{equation}
B_{\mathcal{F}}(u, e) = \begin{cases}
-\mathcal{F}_{u \triangleleft e} &\text{if node}\,\, u\,\, \text{is on the tail of edge}\,\, e;\\
\mathcal{F}_{u \triangleleft e} &\text{if node}\,\, u\,\, \text{is on the head of edge}\,\, e;\\
0 &\text{elsewhere.}
\end{cases}
\end{equation}
with $u=1, \dots, V$, and $e=1, \ldots, E$, such that $B^1_{\mathcal{F}} = \delta^T$.
\begin{definition}[Sheaf laplacian]
The \textit{sheaf laplacian} is a linear operator $L_{\mathcal{F}} = \delta^T\delta: C^0(G,\mathcal{F}) \rightarrow C^0(G,\mathcal{F})$ such that:
    \begin{gather}
        (L_{\mathcal{F}})_{uu} = \sum_{e : u\triangleleft e} \mathcal{F}_{u \triangleleft e}^T\mathcal{F}_{u \triangleleft e} \\ 
        (L_{\mathcal{F}})_{uv} =  -\mathcal{F}_{u \triangleleft (u,v)}^T\mathcal{F}_{v \triangleleft (u,v)}.
    \end{gather}
\end{definition}
The \textit{sheaf Laplacian} is a matrix that captures both the combinatorial structure of the graph and the geometric properties imparted by the sheaf. Its block structure reflects the sparsity pattern of the underlying graph, while the interaction between stalks is encoded through the restriction maps provided by the sheaf. This structure subsumes the classic graph as a specific case called constant sheaf, where each stalk is $\mathbb{R}$ and each map is the identity map. This motivates the interest towards spectral sheaf theory as a generalization of graph spectral theory. The most relevant result in this sense connects the space of global sections to the spectrum of the sheaf Laplacian: 
\begin{theorem}[Hodge Theorem]
For \(\mathcal{F}\) a sheaf on a graph \(G\) as above,
\[
H^0(G; \mathcal{F}) = \ker L_{\mathcal{F}}.
\]
\end{theorem}

Hence, for a sheaf \((G, \mathcal{F})\), where \(G = (V, E)\) is a graph and \(\mathcal{F}\) is a sheaf over \(G\), the quadratic form associated with the sheaf Laplacian \(L_{\mathcal{F}}\), given by
\[
\mathbf{x}^T L_{\mathcal{F}} \mathbf{x} = \langle \mathbf{x}, L_{\mathcal{F}} \mathbf{x} \rangle,
\]
where \(\mathbf{x}\) is a \(0\)-cochain, serves as a measure of the \textit{smoothness} of the signal \(\mathbf{x}\) over \((G, \mathcal{F})\). Specifically, the term \(\mathbf{x}^T L_{\mathcal{F}} \mathbf{x}\) quantifies the extent to which \(\mathbf{x}\) is consistent with the restriction maps defined by the sheaf: 

\begin{itemize}
    \item If \(\mathbf{x}^T L_{\mathcal{F}} \mathbf{x}\) is small, it indicates that \(\mathbf{x}\) is nearly a \textit{global section} of the sheaf, meaning that the values assigned by \(\mathbf{x}\) to the stalks align well under the restriction maps along the edges of \(G\).
    \item Conversely, a large value of \(\mathbf{x}^T L_{\mathcal{F}} \mathbf{x}\) suggests significant discrepancies between the values assigned to adjacent nodes, as propagated through the restriction maps.
\end{itemize}

Being a positive semi-definite matrix by construction, the sheaf Laplacian induces a \textit{semi-inner product} \(\langle \mathbf{x}, L_{\mathcal{F}} \mathbf{x} \rangle\) that, in turn, defines a notion of semi-distance between the signal \(\mathbf{x}\) and the space of global sections of the sheaf \((G, \mathcal{F})\). This semi-distance reflects how far the signal is from satisfying the consistency conditions imposed by the sheaf's structure.

\section{The global problem}

In classic graph representation learning, the  notion of consistency embedded in the graph Laplacian is the starting point to define learning algorithms aimed to minimize the total variation of the observed signals $\mathbf{X} \in \mathbb{R}^{|V| \times N}$, where $N$ is the number of snapshots, over the graph: 
\begin{equation}
\underset{L \in \mathcal{L}}{\mathrm{min}} \ \mathrm{tr}(\mathbf{X}^T L \mathbf{X}) + f(L)
\end{equation}
where $f(L)$ is a regularizer encouraging some desiderata for the retrieved network in terms of connectivity and sparsity, and $\mathcal{L}$ is a proper convex cone of graph Laplacians. 
Inspired by this approach and by the work in \cite{pmlr-v51-kalofolias16}, the authors in \cite{8683709} proposed a Symmetric Positive Semi Definite (SPSD) programming formulation for the sheaf Laplacian learning problem based on observed 0-co-chains $\mathbf{X} \in \mathbb{R}^{d|V| \times N}$, assuming that all node stalks have the same dimension $d$.

This formulation inherits the elegance and the structuring of SPSD programming and graph learning, but the graph sheaf Laplacian operator does not provide enough information to extract the isomorphism class of the sheaf itself. In other words this means that learning the sheaf Laplacian does not allow us to specify $(G,\mathcal{F})$ well enough in terms of the restriction maps acting along each edge.

Our proposal aims to estimate directly the restriction maps $\mathcal{F}_{u \triangleleft e}$ and $\mathcal{F}_{v \triangleleft e}$ associated to each edge, by rewriting the total variation as: 

\begin{align}
\operatorname{tr}\left( \mathbf{X}^T L_{\mathcal{F}} \mathbf{X} \right) &= \operatorname{tr}\left( \mathbf{X}^T B_{\mathcal{F}}
B^T_{\mathcal{F}} \mathbf{X} \right) \\
&= \sum_{e \in E} \lVert \mathcal{F}_{u \triangleleft e} \mathbf{X}_u - \mathcal{F}_{v \triangleleft e} \mathbf{X}_v \rVert_F^2
\end{align}


Extending the approach proposed in \cite{chepuri2016learningsparsegraphssmoothness}, we formulate the inference problem as the minimization of the total variation with respect to all the restriction maps and to a set of binary decision variables, one for each possible edge, used to decide whether an edge is present or not. The cardinality $E_0$ of the edge set is assumed to be a priori known.
In practice this number is not known and it has to be computed through cross validation. We also need to impose a constraint on the set of feasible restriction maps ${\mathcal F}$, to avoid the trivial solution. 
\footnotesize
\[
\begin{aligned}
\begin{split}
\begin{cases}
\underset{\{\mathcal{F}_{u \triangleleft e},\mathcal{F}_{v \triangleleft e}, a_e\}_{e \in \mathcal{E}}}{\mathrm{min}} & \sum_{e \in \mathcal{E}} a_e ||\mathcal{F}_{u \triangleleft e}\mathbf{X}_u - \mathcal{F}_{v \triangleleft e}\mathbf{X}_v||^2 \\
& ||a||_0 = E_0;  \mathcal{F}_{u \triangleleft e}\,\, {\rm{and}}\,\, \mathcal{F}_{v \triangleleft e} \in \mathcal{F}\\
& a_e \in \{0,1\}, \forall e \in \mathcal{E}
\end{cases}
\end{split}
\end{aligned}
\]
\normalsize
Despite being a combinatorial problem, its solution can be easily found in a closed form. In particular: 
    \begin{itemize}
        \item Compute the solution for the local problem $\mathcal{P}_e$ on each of the possible connections $e \in \mathcal{E}$;
        \item Sort the set $\mathcal{E}$ according to the value of the objective function for $\mathcal{P}_e$;
        \item Set $a_e = 1$ for the first $E_0$ retrieved edges, $a_e = 0$ for all the others.
    \end{itemize}
With this approach, the space complexity of each local problem $\mathcal{P}_e$ is $\mathcal{O}(d^2)$ if we assume the same dimension $d$ on each stalk, while the required full enumeration of all the connections makes its overall complexity $\mathcal{O}(V^2d^2)$, but the decomposition may enable parallelization and distribution of the workload.

Before solving the local problem, we apply a denoising step to the observed data by representing the data set over a known dictionary $\mathbf{D}$ and searching for the sparse set of coefficient $\mathbf{S}_u$, on each node $u$, that minimizes the following objective function
\begin{equation}
\label{sparse representation}
    \underset{\mathbf{S_u} }{\mathrm{min}} \ \sum_{u=1}^{V}\|\mathbf{X}_u-\mathbf{D}\mathbf{S}_u\|_F^2+\alpha\,\sum_{u=1}^{V} \|\mathbf{S}_u\|_{2,1},
\end{equation}
where we use the $\ell_{2,1}$-norm to promote block sparsity along the rows of the matrices $\mathbf{S}_u$. We denote by $\mathbf{D}_u \mathbf{S}_u$ the compact representation found on each node $u$. In general, after this denoising step, the signal on each node lives in a subspace of dimension $d_u$ that may be different for each node.

\section{The local problems}
The feasible set to be used in the search of the restriction maps in the global problem can be chosen in different ways. 
A possibility that is theoretically plausible and gives rise to a simple solution is that of assuming the restriction maps to be orthonormal. In this way, the restriction maps represent isometries. The local problem, to be solved for each potential edge is then: 
\[
\begin{aligned}
\begin{split}
\begin{cases}
\underset{\{\mathcal{F}_{u \triangleleft e},\mathcal{F}_{v \triangleleft e}\}}{\mathrm{min}} &  ||\mathcal{F}_{u \triangleleft e}\mathbf{D}_u \mathbf{S}_u - \mathcal{F}_{v \triangleleft e}\mathbf{D}_v \mathbf{S}_v||^2 \\
& \mathcal{F}_{u \triangleleft e}^T\mathcal{F}_{u \triangleleft e} = \mathbb{I} \\
& \mathcal{F}_{v \triangleleft e}^T\mathcal{F}_{v \triangleleft e} = \mathbb{I}
\end{cases}
\end{split}
\end{aligned}
\]
Clearly, the solution can be found only up to a rotation matrix. Hence we can equivalently recast the problem  by searching for a single rotation matrix 
$\mathcal{F}_{u \triangleleft e}$, while
setting the other matrix to be the identity matrix, i.e. $\mathcal{F}_{v \triangleleft e} = \mathbb{I}$. The local problem is then:
\[
\begin{aligned}
\begin{split}
\begin{cases}
\underset{\mathcal{F}_{u \triangleleft e}}{\mathrm{min}} &  ||\mathcal{F}_{u \triangleleft e}\mathbf{D}_u \mathbf{S}_u - \mathbf{D}_v \mathbf{S}_v||^2 \\
& \mathcal{F}_{u \triangleleft e}^T\mathcal{F}_{u \triangleleft e} = \mathbb{I} 
\end{cases}
\end{split}
\end{aligned}
\]
This problem can be solved as an orthogonal Procrustes problem, so that each local problem can be solved in closed form solutions via singular value decomposition. 

More specifically, expanding the squared norm, we get: 

\small
\[
||\mathcal{F}_{u \triangleleft e}\mathbf{D}_u\mathbf{S}_u - \mathbf{D}_v\mathbf{S}_v||^2 = 
\]
\[
\mathrm{tr}\{(\mathcal{F}_{u \triangleleft e}\mathbf{D}_u\mathbf{S}_u - \mathbf{D}_v\mathbf{S}_v)(\mathcal{F}_{u \triangleleft e}\mathbf{D}_u\mathbf{S}_u - \mathbf{D}_v\mathbf{S}_v)^T\}
\]
\[
= \mathrm{tr}\{\mathcal{F}_{u \triangleleft e}\mathbf{D}_u\mathbf{S}_u\mathbf{S}_u^T\mathbf{D}_u^T\mathcal{F}_{u \triangleleft e}^T + \mathbf{D}_v\mathbf{S}_v\mathbf{S}_v^T\mathbf{D}_v^T - 2\mathcal{F}_{u \triangleleft e}\mathbf{D}_u\mathbf{S}_u\mathbf{S}_v^T\mathbf{D}_v^T\}
\]
\[
= \mathrm{tr}\{\underbrace{\mathcal{F}_{u \triangleleft e}^T\mathcal{F}_{u \triangleleft e}}_{\mathbb{I}}\mathbf{D}_u\mathbf{S}_u\mathbf{S}_u^T\mathbf{D}_u^T + \mathbf{D}_v\mathbf{S}_v\mathbf{S}_v^T\mathbf{D}_v^T - 2\mathcal{F}_{u \triangleleft e}\mathbf{D}_u\mathbf{S}_u\mathbf{S}_v^T\mathbf{D}_v^T\}.
\]
\normalsize
Considering only the term in which $\mathcal{F}_{u \triangleleft e}$ does appear, the original problem can be rewritten as
\[
\begin{aligned}
\begin{split}
\begin{cases}
\underset{\mathcal{F}_{u \triangleleft e}}{\mathrm{max}} &  \mathrm{tr}\{\mathcal{F}_{u \triangleleft e}\mathbf{D}_u\mathbf{S}_u\mathbf{S}_v^T\mathbf{D}_v^T   \} \\
& \mathcal{F}_{u \triangleleft e}^T\mathcal{F}_{u \triangleleft e} = \mathbb{I} 
\end{cases}
\end{split}
\end{aligned}
\]
The inner product denotes an empirical cross-covariance $\hat{\mathbf{C}}_{uv} = \mathbf{S}_u\mathbf{S}_v^T/N$. This implies, as to be expected, that an alignment between the signals living over two nodes $u$ and $v$ is possible, only if the signals observed over the two nodes exhibit some kind of correlation. 
Hence, under the assumption of a non-trivial statistical dependency structure, the trace to be maximized is:
\[
\mathrm{tr}\{\mathcal{F}_{u \triangleleft e}\mathbf{D}_u\hat{\mathbf{C}}_{uv}\mathbf{D}_v^T\} 
= 
\]
\[
\mathrm{tr}\{\mathcal{F}_{u \triangleleft e}\mathbf{U}\Sigma_{uv}\mathbf{V}^T\} 
= 
\]
\[
\mathrm{tr}\{\underbrace{\mathbf{V}^T\mathcal{F}_{u \triangleleft e}\mathbf{U}}_{\mathbf{Z}}\Sigma_{uv}\} 
= \mathrm{tr}\{\mathbf{Z}\Sigma_{uv}\}.
\]
where we have denoted by $\mathbf{U}\Sigma_{uv}\mathbf{V}^T$ the Singular Value Decomposition (SVD) of $\mathbf{D}_u\hat{\mathbf{C}}_{uv}\mathbf{D}_v^T$.

The problem 
\[
\begin{aligned}
\begin{split}
\begin{cases}
\underset{\mathbf{Z}}{\mathrm{max}} &  \mathrm{tr}\{\mathbf{Z}\Sigma_{uv}\}
 \\
& \mathbf{Z}^T\mathbf{Z} = \mathbb{I} 
\end{cases}
\end{split}
\end{aligned}
\]
is trivially solved by setting $\mathbf{Z} = \mathbb{I}$, which implies, given the chain of equalities, that the original problem is solved by setting $\mathcal{F}_{u \triangleleft e} = \mathbf{V}\mathbf{U}^T$. The resulting maximum trace is then
\begin{equation}
    \mathrm{tr}\{\mathbf{Z}\Sigma_{uv}\}=\sum_{i=1}^{r_{uv}} \sigma_{uv}(i),
\end{equation}
where $\sigma_{uv}(i)$, $i=1, \ldots r_{uv}$, are the singular values of $\Sigma_{uv}$ and $r_{uv}$ is the rank of $\Sigma_{uv}$. It is also easy to check that, if the column of the dictionary $\mathbf{D}$ are orthonormal, the singular values of  $\mathbf{D}_u\hat{\mathbf{C}}_{uv}\mathbf{D}_v^T$ coincide with the singular values of $\hat{\mathbf{C}}_{uv}$. This shows that, after alignment,  the distance between each pair of nodes depends only on the covariance between the signals living on the edge connecting the two nodes, quite interestingly, on the dimensions of the subspaces spanned by the local dictionaries $\mathbf{D}_u$ and $\mathbf{D}_v$, but not on their structure.
\section{Numerical results}

\textbf{Total variation on graphs and on sheaves} As a first experiment, we wish to compare the total variation achievable with a conventional graph inference algorithm, based on the conventional distance between signals pertaining to different nodes of the graph, with the total variation obtained with our approach, using the same number of edges in both cases.
To carry out this comparison, we generated the synthetic data  as follows.
First, we selected a standard orthonormal basis $\mathbf{D}$ of $\mathbb{R}^d$, which serves as a global dictionary. Then, for each node, we randomly defined a linear subspace of $\mathbb{R}^d$ by sampling a subset of elements from $\mathbf{D}$, with a random dimensionality. Then we generated the coefficients $\mathbf{S}_u$ of the representation in each node as Gaussian random variables and we impressed some correlation between coefficients pertaining to different nodes. Finally, we added Gaussian white noise $\mathbf{N}_v$ in each observation. The resulting signal observed in each node has then the form $\mathbf{Y}_v = \mathbf{D}_v \mathbf{S}_v + \mathbf{N}_v$, where $\mathbf{D}_v$ is the local basis at node $v$. We assumed the same signal-to-noise ratio (SNR) in each node, for the sake of simplicity. 
\begin{figure}[h!]
    \centering
    \includegraphics[width=0.8\linewidth]{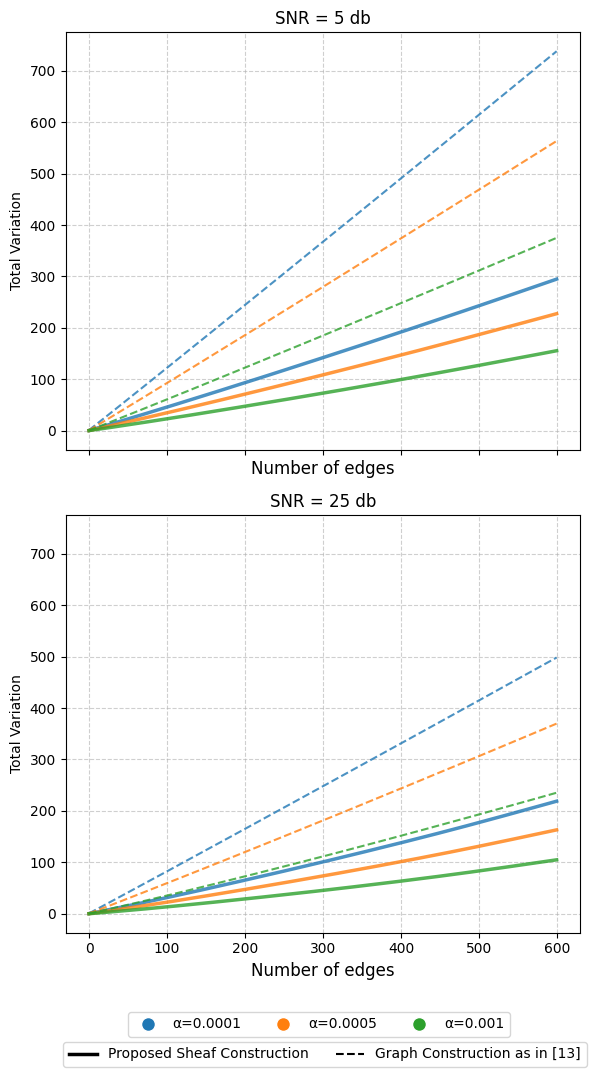}
    \caption{Total variation for the proposed construction and a graph construction for two SNRs with respect to different values for $\alpha$ and number of edges}
    \label{fig:total variation}
\end{figure}
In Fig. \ref{fig:total variation} we show the comparison between the total variation obtained from a conventional approach (dashed line) and with our approach (solid line), as a function of the number of edges of the inferred graph, the coefficient $\alpha$ used in \eqref{sparse representation} to promote a sparse representation in each node, and for two values of the SNR. We can clearly see how the inferred sheaf Laplacian makes the data representation smoother with respect to a conventional method, thanks to the optimization of the alignment matrices $\mathcal{F}_{u \triangleleft e}$. The method is also quite robust with respect to additive noise.

\begin{figure}
    \centering
    \includegraphics[width=0.7\linewidth]{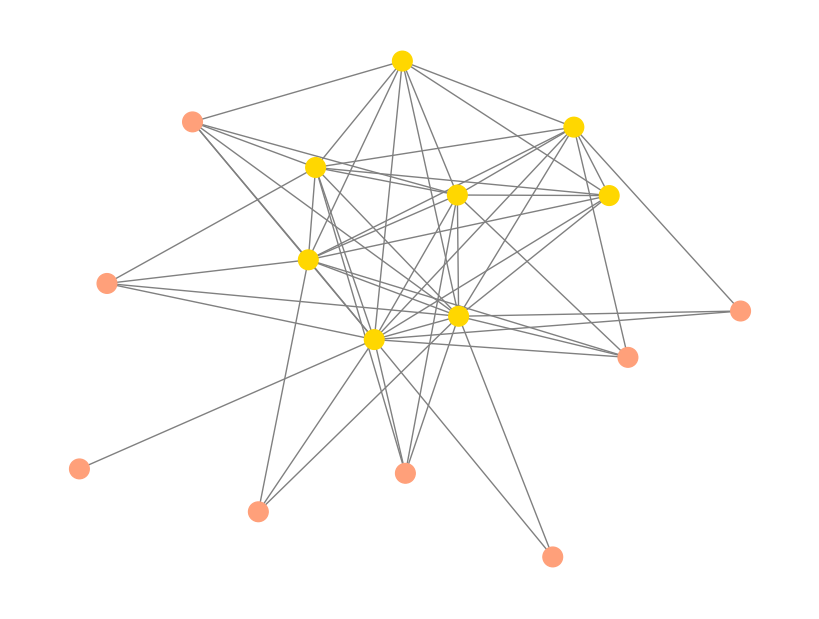}
    \caption{Graph obtained via hierarchical clustering without alignment ($E_0 = 57$ is minimum for connection)}
    \label{fig:graph before alignment}
\end{figure}

\begin{figure}
    \centering
    \includegraphics[width=0.7\linewidth]{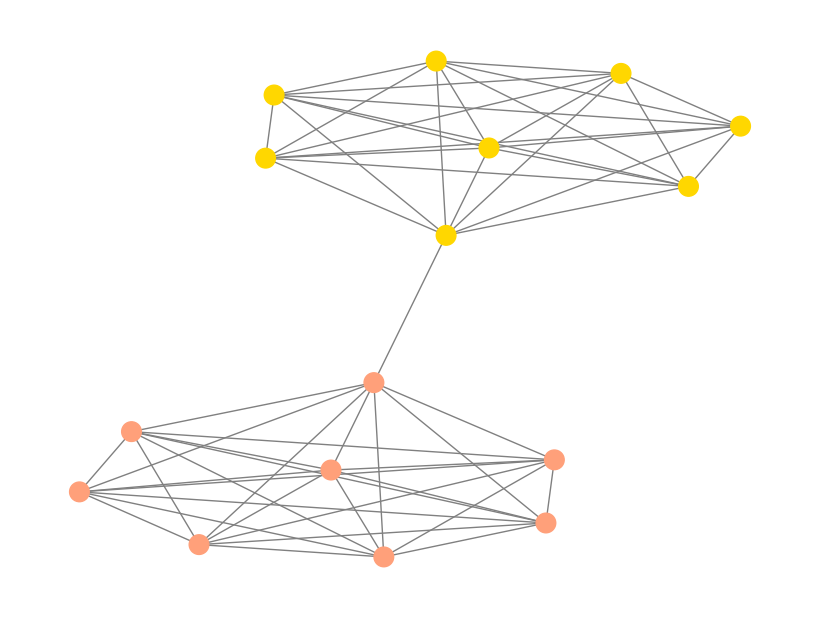}
    \caption{Graph obtained via hierarchical clustering with alignment ($E_0 = 56$ is minimum for connection)}
    \label{fig:graph after alignment}
\end{figure}

\textbf{Clustering and structured connnectomy} A desirable property when constructing a cellular sheaf out of the observed signals is to have some spectral qualities reflecting how local and global sections can arise within the considered sheaf.
In our theoretical findings, we showed that the distance between signals living over each pair of nodes depends on the cross-correlation between the data living in different nodes and on the dimensions of the subspaces where the signals in each node lie, but not directly on the local dictionaries. To test this interesting property, we generated data over
$16$ nodes, assigning local subspace dimensions such that half of them admits a sparse representation over $\mathbb{R}^{10}$ and the other half admits a sparse representation over $\mathbb{R}^{40}$, where $\mathbb{R}^{64}$ is the ambient stalk. The dictionaries used in each node are generated at random. In Figs. \ref{fig:graph before alignment} and \ref{fig:graph after alignment}, we draw two examples of graphs inferred from the same data, using, respectively,  a conventional approach that looks at the distance between the subspaces pertaining to each pair of nodes, and the distance obtained after alignment using our method. Comparing the two figures, we can  observe how our method is able to build a graph with two clear clusters, evidencing the similarity between nodes characterized by the same subspace dimension. Using a conventional distance, no clustering property is evidenced. Furthermore, from Fig. \ref{fig:graph after alignment} we notice how, within each cluster, the nodes tend to be all connected, even though their dictionaries are different.

\section{Conclusions}
In this work, we introduced a novel framework for infer the topology of a graph, together with the restriction maps, in order to minimize the total variation of the observed data set. The method is much more numerically efficient than alternative approaches based on semidefinite programming. We assessed the effectiveness of our method comparing it with a typical graph inference based on conventional (i.e., without alignment) distance measures between pairs of nodes. We plan to extend our work to make it more scalable when working out a joint sheaf learning and signal denoising task, generalize it to cellular sheaves defined over higher order topologies leveraging discrete Hodge theory and the interplay between higher order cochains signals, and to refine the formulation of the initial global problem to incorporate additional information on the inferred network.



\normalsize

\bibliographystyle{unsrt}
\bibliography{bibliography}


\end{document}